\documentclass[showpacs,preprintnumbers]{revtex4}
\usepackage{epsfig}

\newcommand{\beq}{\begin{equation}}
\newcommand{\eeq}{\end{equation}}

\newcommand{\etal}{{\rm et al.}}

\newcommand{\sigv}{\langle\sigma v\rangle}

\def\erf{\mathop{\rm erf}}

\newcommand{\app}[3]{Astropart.\ Phys.\ {\bf #1}, #3 (#2)}

\newcommand{\hepph}[1]{{\tt hep-ph/#1}}
\newcommand{\astroph}[1]{{\tt astro-ph/#1}}
\newcommand{\prep}[3]{Phys.\ Rep.\ {\bf #1}, #3 (#2)}
\newcommand{\plb}[3]{Phys.\ Lett.\ B\ {\bf #1}, #3 (#2)}
\newcommand{\npb}[3]{Nucl.\ Phys.\ B\ {\bf #1}, #3 (#2)}

\renewcommand{\apj}[3]{Astrophys.\ J.\ {\bf #1}, #3 (#2)}
\newcommand{\apjs}[3]{Astrophys.\ J.\ Suppl.\ Ser.\ {\bf #1}, #3 (#2)}

\newcommand{\aeta}[3]{Astron.\ {\&}\ Astrophys.\ {\bf #1}, #3 (#2)}
\newcommand{\mnras}[3]{Mon.~Not.~R.~Astron.~Soc.\ {\bf #1}, #3 (#2)}
\renewcommand{\prl}[3]{Phys.\ Rev.\ Lett. {\bf #1}, #3 (#2)}
\renewcommand{\prd}[3]{Phys.\ Rev.\ D\ {\bf #1}, #3 (#2)}

\begin{document}

\title{Diffuse inverse Compton and synchrotron emission\\ from dark matter
annihilations in galactic satellites}

\author{E.~A.~Baltz}
\email{eabaltz@slac.stanford.edu}
\affiliation{KIPAC, MS 29, P.O.\ Box 20450, Stanford University, Stanford, CA
94309}
\author{L.~Wai}
\email{wai@slac.stanford.edu}
\affiliation{SLAC, MS 98, 2575 Sand Hill Road, Menlo Park, CA 94025}

\date{\today}

\begin{abstract}
Annihilating dark matter particles produce roughly as much power in electrons
and positrons as in gamma ray photons.  The charged particles lose essentially
all of their energy to inverse Compton and synchrotron processes in the
galactic environment.  We discuss the diffuse signature of dark matter
annihilations in satellites of the Milky Way (which may be optically dark with
few or no stars), providing a tail of emission trailing the satellite in its
orbit.  Inverse Compton processes provide X-rays and gamma rays, and
synchrotron emission at radio wavelengths might be seen.  We discuss the
possibility of detecting these signals with current and future observations, in
particular EGRET and GLAST for the gamma rays.
\end{abstract}


\maketitle

\section{Introduction}
It is almost universally accepted that most of the matter in the universe is
non-baryonic.  This dark matter is the chief constituent of gravitationally
bound objects from dwarf galaxy scales and larger.  Identifying the nature of
dark matter is one of the most important problems in astrophysics, cosmology,
and particle physics.

Perhaps the best motivated candidate for cold dark matter is the lightest of
the so--called neutralinos arising in supersymmetric extensions to the standard
model \cite{jkg96}.  These are the spin-1/2 Majorana fermion counterparts of
the neutral gauge and Higgs bosons, and are expected to have masses at the weak
scale (of order 100 GeV).  This scale is intriguing as the relic density of a
stable particle with this mass and corresponding cross section turns out to be
of order the critical density, as we observe the matter density to be today.
Any stable particles with weak scale masses could thus naturally account for
the dark matter.  In this paper, we will focus on supersymmetry, but our
conclusions are fairly generic to dark matter candidates at the TeV scale.

We outline a new signature of annihilating dark matter in satellites of the
Milky Way galaxy.  Prospects for detecting high energy photons as dark matter
annihilation products, primarily from the $\pi^0$ decays that are generic to
hadronization processes, have been discussed for many years (for a sample see
Ref.~\cite{gammas}).  Necessarily coming with these photons are high energy
electrons and positrons from the analogous $\pi^\pm$ decay chains.  Charged
particles suffer complicated motions in the galactic magnetic field, and
furthermore they lose energy to synchrotron and inverse Compton processes.
Searching for the synchrotron emission from the galactic center \cite{syncGC}
and from galactic satellites \cite{blasi} has been discussed previously, though
neglecting the diffusion of the charged particles.  We will show that the
inverse Compton emission, extended over a large area from the charged particle
annihilation products may be observable for some models of particle dark matter
and of galactic satellites.

\section{Supersymmetric candidates}

\subsection{Particle physics model}

In the Minimal Supersymmetric Standard Model (MSSM) the lightest of the
superpartners (LSP) is often the lightest neutralino.  The latter is a
superposition of the superpartners of the neutral gauge and Higgs bosons,
\begin{equation}
\tilde{\chi}^0_1 =
N_{11} \tilde{B} + N_{12} \tilde{W}^3 +
N_{13} \tilde{H}^0_1 + N_{14} \tilde{H}^0_2.
\end{equation}
With R-parity conserved, this lightest superpartner is stable.  For significant
regions of the MSSM parameter space, the relic density of the stable neutralino
is of the order $\Omega_\chi h^2\sim0.1$, thus constituting an important (and
perhaps exclusive) part of the cold dark matter.  Note that $\Omega_{\chi}$ is
the neutralino density in units of the critical density and $h$ is the present
Hubble constant in units of $100$ km s$^{-1}$ Mpc$^{-1}$.  Current
observations, included those of the WMAP satellite and the Sloan Digital Sky
Survey (we take the WMAP values), favor $h=0.71^{+0.04}_{-0.03}$ and a matter
density $\Omega_{M}h^2 = 0.135^{+0.008}_{-0.009}$, of which baryons contribute
a small amount $\Omega_Bh^2=0.0224\pm 0.0009$ \cite{wmap,sdss}.  If we assume
that neutralinos are the only constituent of matter (neutrinos are a small
contribution, $\Omega_\nu h^2<0.0076$), we can then infer $\Omega_\chi
h^2=0.113^{+0.008}_{-0.009}$.  We will apply a generous 3$\sigma$ constraint on
the relic abundance from the WMAP data as follows: $0.086<\Omega_\chi
h^2<0.137$.

Using the DarkSUSY code \cite{darksusy}, we have explored the supersymmetric
parameter space in both a phenomenological MSSM
\cite{bg,coann,jephd,bub,neutrate,other_db,be99,sfermioncoann} and in minimal
supergravity (using the ISAJET code \cite{isajet}).  Each model is subjected to
current accelerator constraints on masses of superpartners and Higgs bosons
\cite{pdg2002,lepsusy} and on the $b\rightarrow s\gamma$ branching ratio
\cite{cleo}.  Crucial for studies of dark matter, the relic abundance of
neutralinos $\Omega_{\chi} h^2$ is calculated based on
Refs.~\cite{coann,GondoloGelmini,sfermioncoann}.

\subsection{Photon and electron spectra of annihilations}

\begin{figure}
\epsfig{file=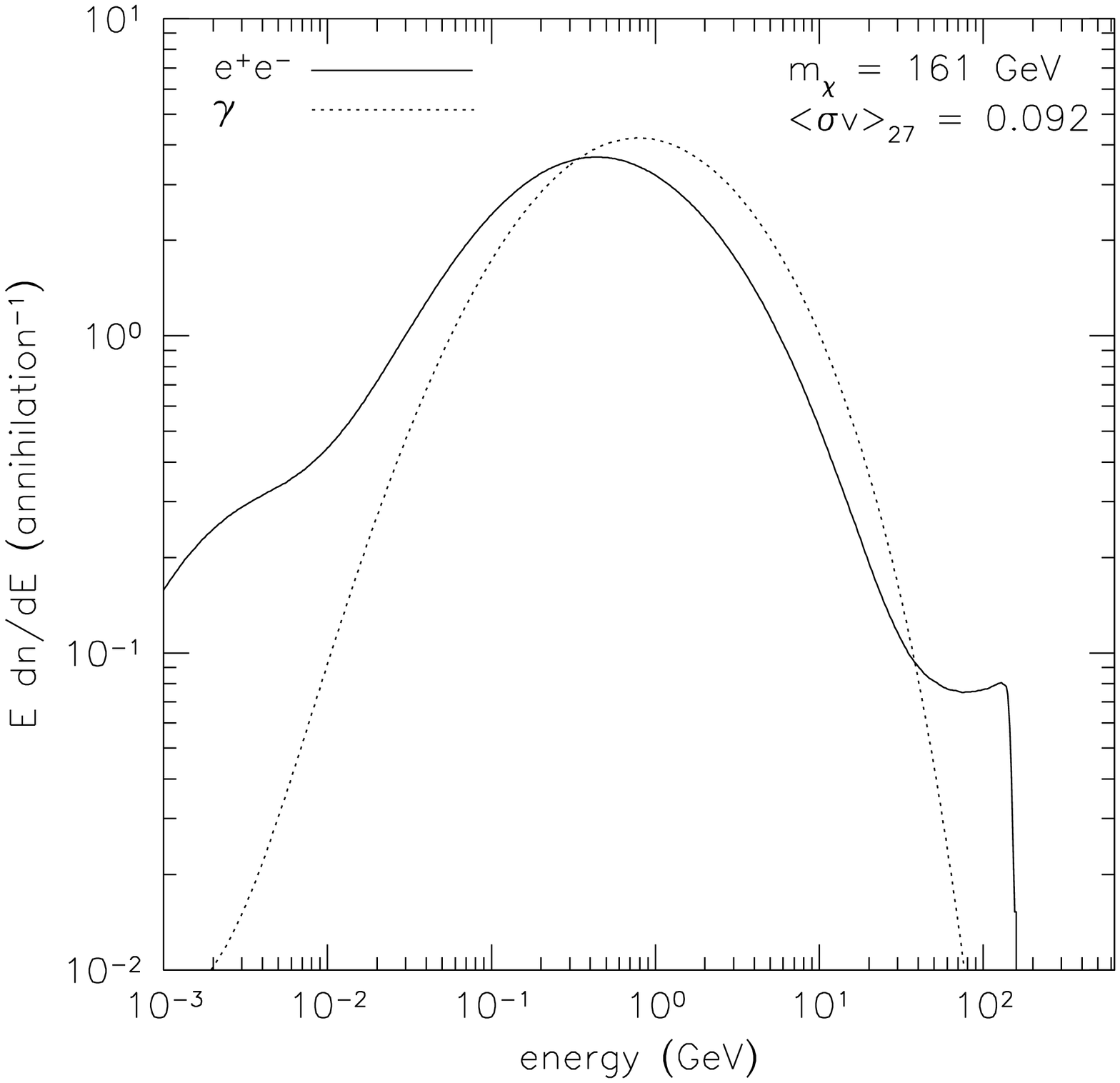,width=0.49\textwidth}
\epsfig{file=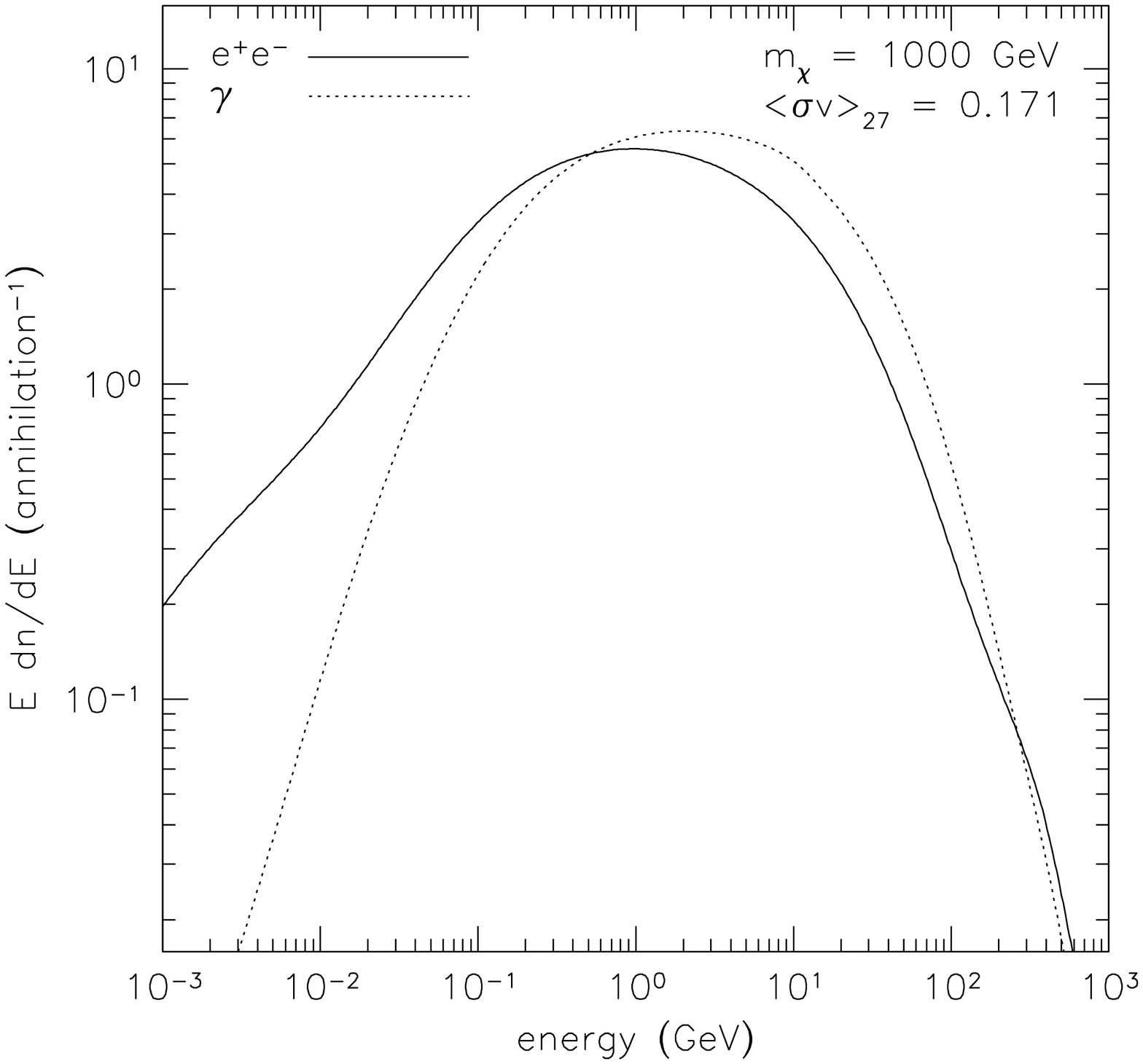,width=0.49\textwidth}
\caption{Spectra of photons and leptons from annihilations in two
characteristic models.  The solid curves depict the combined electron and
positron spectra, and the dotted curves show the photon spectra.  Both of these
models have acceptable values of $\Omega_\chi h^2$.  The model on the left has
an appreciable Higgsino fraction, and thus annihilates efficiently into $W$
pairs.  The direct decays $W^\pm\rightarrow e^\pm\nu$ are responsible for the
shelf in the lepton spectrum at energies right below maximum.  The model on the
right is a fairly pure gaugino, thus the final state is nearly always
$q\overline{q}$, and the spectrum is featureless.}
\label{fig:inputspec}
\end{figure}

The only detailed information about models we require for this study (apart
from the relic density constraint) is the annihilation cross section at
non-relativistic velocities $\sigv$ (accounting for the factor of two due to
identical particles in the initial state) and the spectrum of annihilation
products, in particular the photons, electrons and positrons.  The photons are
primarily from the $\pi^0$ decays associated with hadronic final states,
whereas the leptons can be from the associated $\pi^\pm$ decays, or more
directly from massive gauge boson decays.  In Fig.~\ref{fig:inputspec} we
illustrate two models in the generic MSSM, one with a typical featureless
spectrum of annihilation products, and the other clearly exhibiting the feature
from $W^\pm$ decays.

\section{Charged particle propagation and diffuse emission}

The calculation of the diffuse emission from the annihilations proceeds in
several steps.  First, the time dependent density of charged particles is
determined according to a diffusion model.  The charged particles are trapped
by the galactic magnetic field, which extends for several kpc from the stellar
disk.  Here we stress that we are concerned only with those galactic satellites
that are currently (or in the recent past) within this ``diffusion zone''.  A
time dependent treatment is necessary because a satellite moving with the
typical galactic velocity of 300 km s$^{-1}$ crosses the diffusion zone in
roughly one diffusion time: both timescales are of order tens of millions of
years.  The second step in the calculation is to calculate the column density
of particles along various lines of sight, as a function of particle energy.
Lastly, the inverse Compton and synchrotron spectra can be calculated from the
particle spectrum under some assumptions about the galactic magnetic field and
radiation fields.

\subsection{Diffusion model}

The propagation of charged particles in the tangled galactic magnetic field can
be modeled as diffusion.  Electrons and positrons lose energy rapidly to
synchrotron radiation, and also to inverse Compton scattering on the cosmic
microwave background (CMB) and on starlight.  We will use the diffusion model
of Ref.~\cite{be99}.  More sophisticated models are available, e.g.\ the
semianalytic treatment of Ref.~\cite{salati} and the fully numerical GALPROP
model \cite{galprop}, but our model is simple to implement, yielding
quantitatively similar results, and affords an intuitive understanding.  We
will make the assumption that the charged particle velocity distribution is
locally isotropic, namely that the particles ``stick'' to the galactic magnetic
field, and that the velocity distribution contains no record of the initial
conditions.  This is important as we consider sources (satellite halos) that
move a significant distance in a diffusion time.  If the velocity distribution
was isotropic in the satellite frame (as could be the case if the satellite had
a significant magnetic field of its own), our results would be different.

The time dependent diffusion--loss equation in homogeneous space for the
density of charged particles as a function of energy $dn/dE$ is the following,
\begin{equation}
\frac{\partial}{\partial t}\frac{dn}{dE}=K(E)\nabla^2
\frac{dn}{dE}+\frac{1}{E_0\tau}\frac{\partial}{\partial E}
\left(E^2\frac{dn}{dE}\right)+Q,
\end{equation}
where $K(E)=K_03^\alpha[1+(E/3\;\rm GeV)^\alpha]$ is the energy dependent
diffusion constant, with $K_0=3\times 10^{27}$ cm$^2$ s$^{-1}$ and
$\alpha=0.6$.  Energy loss due to any electromagnetic process where the
exchanged energy is much less than $m_ec^2$ scales as $E^2$, and is
proportional to the energy density in photons (or equivalently magnetic field).
We consider synchrotron losses dues to a 3 $\mu$G galactic magnetic field (0.2
eV cm$^{-3}$), inverse Compton losses from the CMB (0.3 eV cm$^{-3}$), and
inverse Compton losses due to starlight (0.6 eV cm$^{-3}$) \cite{longair}.  We
thus set $dE/dt=E^2/(E_0\tau)$, with $E_0=1$ GeV, and $\tau=10^{16}$ s.
Lastly, Q is the source function.  Solving instead for $F=E^2\,dn/dE$, we find
\begin{equation}
\left(\frac{1}{E^2}\frac{\partial}{\partial t}-\frac{K(E)}{E^2}\nabla^2
-\frac{1}{E_0\tau}\frac{\partial}{\partial E}\right)F\equiv{\cal L}F=Q.
\end{equation}
We can calculate the Green function for this operator simply by 4-dimensional
Fourier transform in $\vec{x}\rightarrow\vec{k}$ and $t\rightarrow\omega$,
\begin{eqnarray}
{\cal L}G=\delta^3(\vec{x}-\vec{x}')\delta(t-t')\delta(E-E')&\rightarrow& \\
\left[\frac{1}{E^2}\left(-i\omega+K(E)k^2\right)-\frac{1}{E_0\tau}
\frac{\partial}{\partial E}\right]
\tilde{G}&=&\frac{1}{(2\pi)^4}\,e^{i(\omega t'+\vec{k}\cdot\vec{x}')}\,
\delta(E-E').
\end{eqnarray}
We should note that the Green function vanishes for $t'>t$ and for $E'<E$
because time increases monotonically, and energy decreases monotonically.
This equation is easily solved for $E\neq E'$:
\begin{equation}
\tilde{G}=\tilde{G}_0e^{i[\omega\tau (u-u')-K_0\tau k^2(v-v')]},
\end{equation}
with $u=E_0/E$ and $v=3^\alpha u+u^{(1-\alpha)}/(1-\alpha)$.  Hereafter we will
denote the difference $X-X'=\Delta X$ for some variable $X$.  We now apply a
jump condition to fix $\tilde{G}_0$: for $E>E'$, $\tilde{G}=0$.  The
condition is
\begin{equation}
-\frac{1}{E_0\tau}\Delta\tilde{G}=\frac{1}{E_0\tau}\tilde{G}_0=
\frac{1}{(2\pi)^4}e^{i(\omega t'+\vec{k}\cdot\vec{x}')},
\end{equation}
and we thus derive the Fourier transform of the Green function,
\begin{equation}
\tilde{G}=\frac{E_0\tau}{(2\pi)^4}\, e^{i(\omega(\tau\Delta
u+t')+\vec{k}\cdot\vec{x}'-K_0\tau k^2\Delta v)}.
\end{equation}
Inverting the transform, we find the expected behavior that time and energy
propagate in lockstep, yielding the free--space Green function
\begin{equation}
G_{\rm free}=\frac{E_0\tau}{(\pi D^2)^{3/2}}\,e^{-\Delta\vec{x}^2/D^2}\,
\delta(\Delta t-\tau\Delta u),
\end{equation}
defining the effective diffusion length $D^2=4K_0\tau\Delta v=(3.550\;{\rm
kpc})^2\Delta v$.  This scale indicates that particles travel only a few kpc
before losing most of their energy ($\Delta v\sim 1$).

The galactic magnetic field has a limited extent, which can be modeled simply
by defining a diffusion zone at the boundary of which particles freely escape.
Diffusion models typically require that the height of the diffusion zone is
larger than $L=3$ kpc from the disk on both sides, and the radius is at least
20 kpc, and probably larger.  For example the best GALPROP models
\cite{galprop} use $L=4$ kpc and a radius of 30 kpc.  As the radial boundary is
far from the Earth, we can safely neglect it in considering satellites within
10 kpc of us.  We thus model the diffusion zone as an infinite slab of
thickness $2L$, and we take $L=3$ kpc.  We impose the boundary conditions that
the density vanish at $z=\pm L$, which can be effected by a series of image
charges at positions $x_n=x$, $y_n=y$, $z_n=(-1)^nz+2Ln$,
\begin{equation}
G_{2L}(\vec{x},\vec{x}')=\sum_{n=-\infty}^{\infty}(-1)^n\,
G_{\rm free}(\vec{x},\vec{x}'_n).
\end{equation}
The density of charged particles is now derived,
\begin{equation}
F=E^2\frac{dn}{dE}=\int_{R^2} d^2\vec{x}'\int_{-L}^L\!\!\!dz'
\int_{-\infty}^t\!\!\!dt'\int^{\infty}_EdE'\,
G_{2L}(\vec{x},\vec{x}',t,t',E,E')\,Q(\vec{x}',t',E').
\end{equation}

\subsection{Diffuse emission: total power}

For diffuse emission we are interested not in the local density of particles,
but in the column depth of particles,
\begin{equation}
E^2\frac{d\sigma}{dE}=\int d\ell\,E^2\frac{dn}{dE}.
\end{equation}
Placing the observer at the origin, at an angle $\delta$ from the galactic
plane and an angle $\alpha$ from the line $y=0$, we find that for a distance
$\ell$ from the observer,
\begin{eqnarray}
x&=&\ell\,\cos\delta\,\cos\alpha,\\
y&=&\ell\,\cos\delta\,\sin\alpha,\\
z&=&\ell\,\sin\delta.
\end{eqnarray}
For simplicity we define $\hat{x}=\vec{x}/\ell$.  The integral in $\ell$ can be
applied directly to the free Green function, yielding the column depth at
coordinates $(\alpha,\delta)$ due to a source at $\vec{x}'$ (and we truncate at
the edge of the diffusion zone, $\ell_{\rm max}=L/|\sin\delta|$),
\begin{equation}
G_{\rm free}^\sigma=\int_0^{\ell_{\rm max}} \!\!\!d\ell\,G_{\rm free}=
\frac{E_0\tau}{2\pi D^2}e^{[(\hat{x}\cdot\vec{x}')^2-\vec{x}'^2]/D^2}
\left[\erf\left(\frac{\ell_{\rm max}-\hat{x}\cdot\vec{x}'}{D}\right)-
\erf\left(\frac{-\hat{x}\cdot\vec{x}'}{D}\right)\right]
\delta(\Delta t-\tau\Delta u).
\end{equation}
The Green function satisfying the boundary conditions is now simply
\begin{equation}
G_{2L}^\sigma(\alpha,\delta,\vec{x}')=\sum_{n=-\infty}^{\infty}(-1)^n\,
G_{\rm free}^\sigma(\alpha,\delta,\vec{x}'_n),
\end{equation}
and the column density of charged particles is
\begin{equation}
E^2\frac{d\sigma}{dE}=\int_{R^2} d^2\vec{x}'\int_{-L}^L\!\!\!dz'
\int_{-\infty}^t\!\!\!dt'\int^{\infty}_EdE'\,
G_{2L}^\sigma(\alpha,\delta,\vec{x}',t,t',E,E')\,Q(\vec{x}',t',E').
\end{equation}

We consider a toy model for dark matter clump undergoing annihilations.
Assuming a very cuspy profile, most annihilations will occur very close to the
center, and we can assume the clump is a point source.  Taking the clump to be
at position $\vec{X}=(X,Y,Z)$ today $(t=0)$, moving with constant velocity
$\vec{V}$, we can write the source function as
\begin{equation}
Q(\vec{x},t,E)=\Gamma\frac{d\phi}{dE}\delta^3(\vec{x}-\vec{X}-\vec{V}t),
\end{equation}
where $\Gamma$ is the annihilation rate in the clump, and $d\phi/dE$ is the
spectrum of $e^\pm$ per annihilation.  The delta functions simplify matters
greatly, leaving only an integral in energy.  The charged particle density
(both $e^+$ and $e^-$) at the Earth and the column depth are given by
\begin{equation}
E^2\frac{dn}{dE}=\int_E^\infty dE'\,\frac{d\phi}{dE'}
\frac{E_0\Gamma\tau}{(\pi D^2)^{3/2}}\sum_{n=-\infty}^\infty(-1)^n
e^{-\vec{w}_n^2/D^2}\,
\theta\left(L-|w_z|\right),
\end{equation}
\begin{equation}
E^2\frac{d\sigma}{dE}=\int_E^\infty dE'\,\frac{d\phi}{dE'}
\frac{E_0\Gamma\tau}{2\pi D^2}\sum_{n=-\infty}^\infty(-1)^n
e^{[(\hat{x}\cdot\vec{w}_n)^2-\vec{w}_n^2]/D^2}
\left[\erf\left(\frac{\ell_{\rm max}-\hat{x}\cdot\vec{w}_n}{D}\right)-
\erf\left(\frac{-\hat{x}\cdot\vec{w}_n}{D}\right)\right]
\theta\left(L-|w_z|\right).
\end{equation}
The effective position $\vec{w}=\vec{X}-\vec{V}\tau\Delta u$, and
$w_{z,n}=(-1)^n(Z-V_z\tau\Delta u)+2Ln$.  We recover the steady--state solution
by taking $\vec{V}=0$, indicating a source that has been at location $\vec{X}$
for all time.

We observe that looking at lower energies implies looking back in time, to when
the source was in a different position.  This leads to the formation of a
``wake'' of diffuse emission extending from the current position of the source
(which will be a bright, concentrated source of gamma rays) in the direction it
came from.  The spatial signature is not unlike a comet, with a bright
``nucleus'' of gamma rays from the $\pi^0$ decays, and an extended tail of
diffuse emission.

\begin{figure}
\epsfig{file=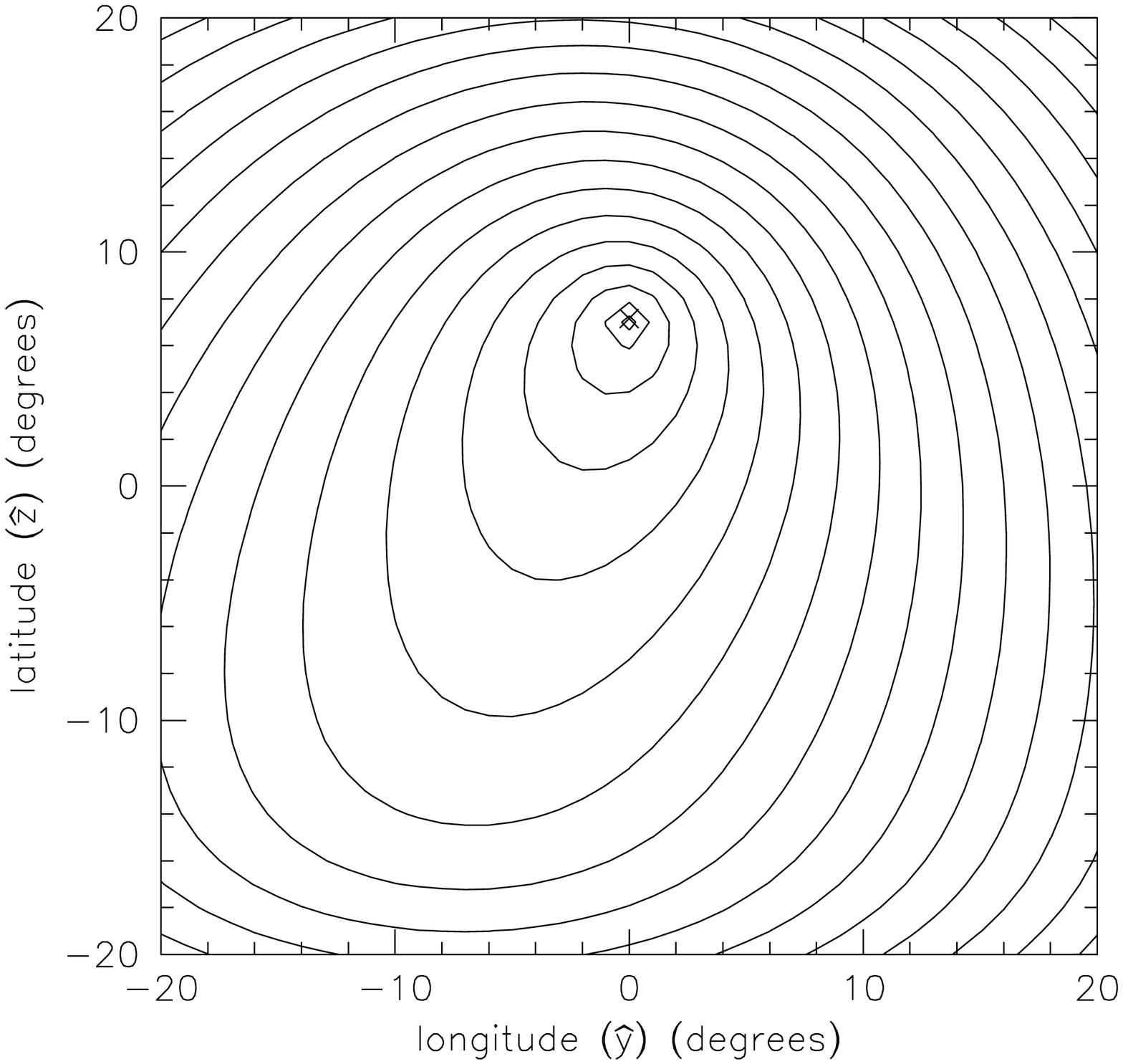,width=0.49\textwidth}
\epsfig{file=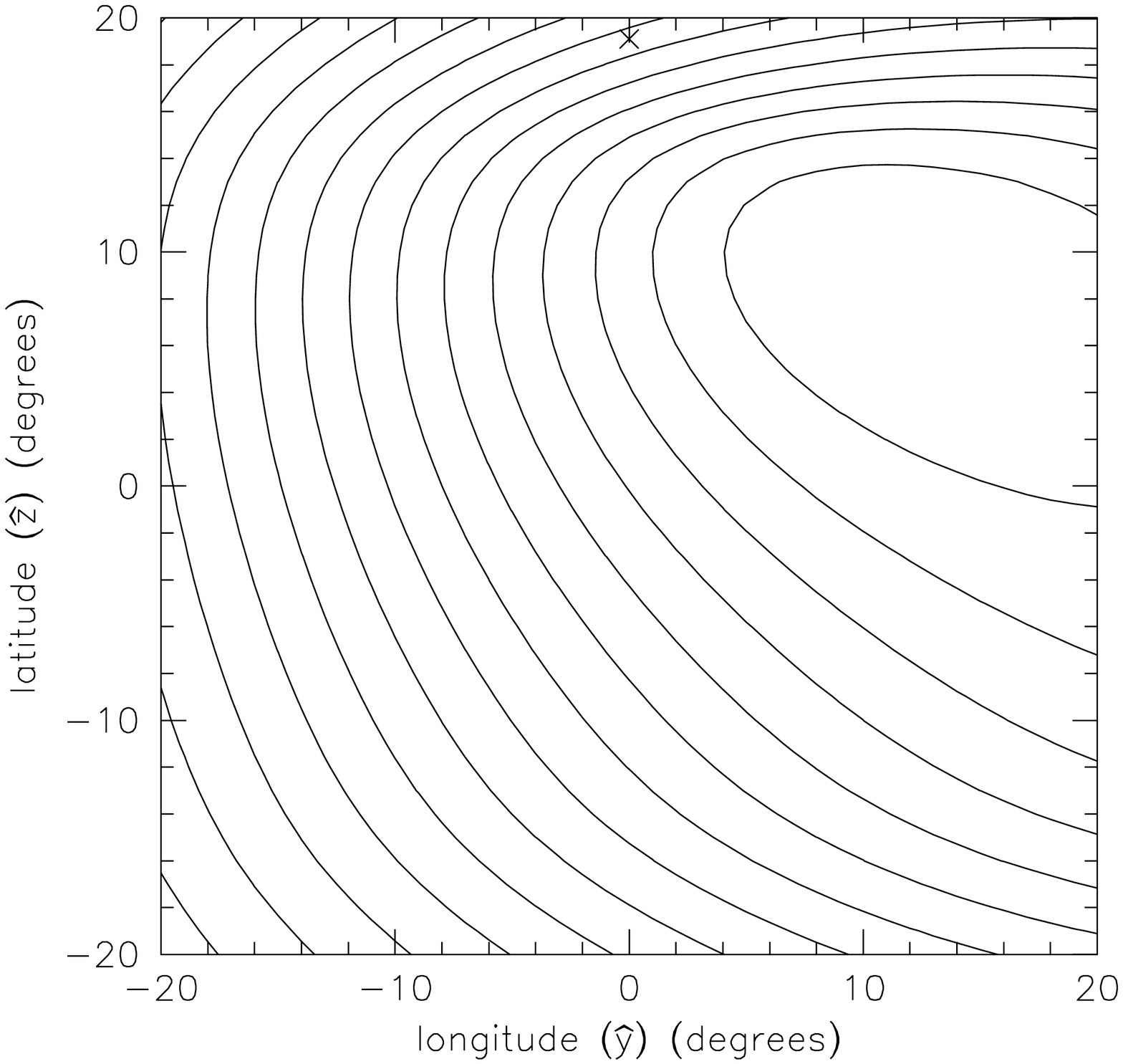,width=0.49\textwidth}
\epsfig{file=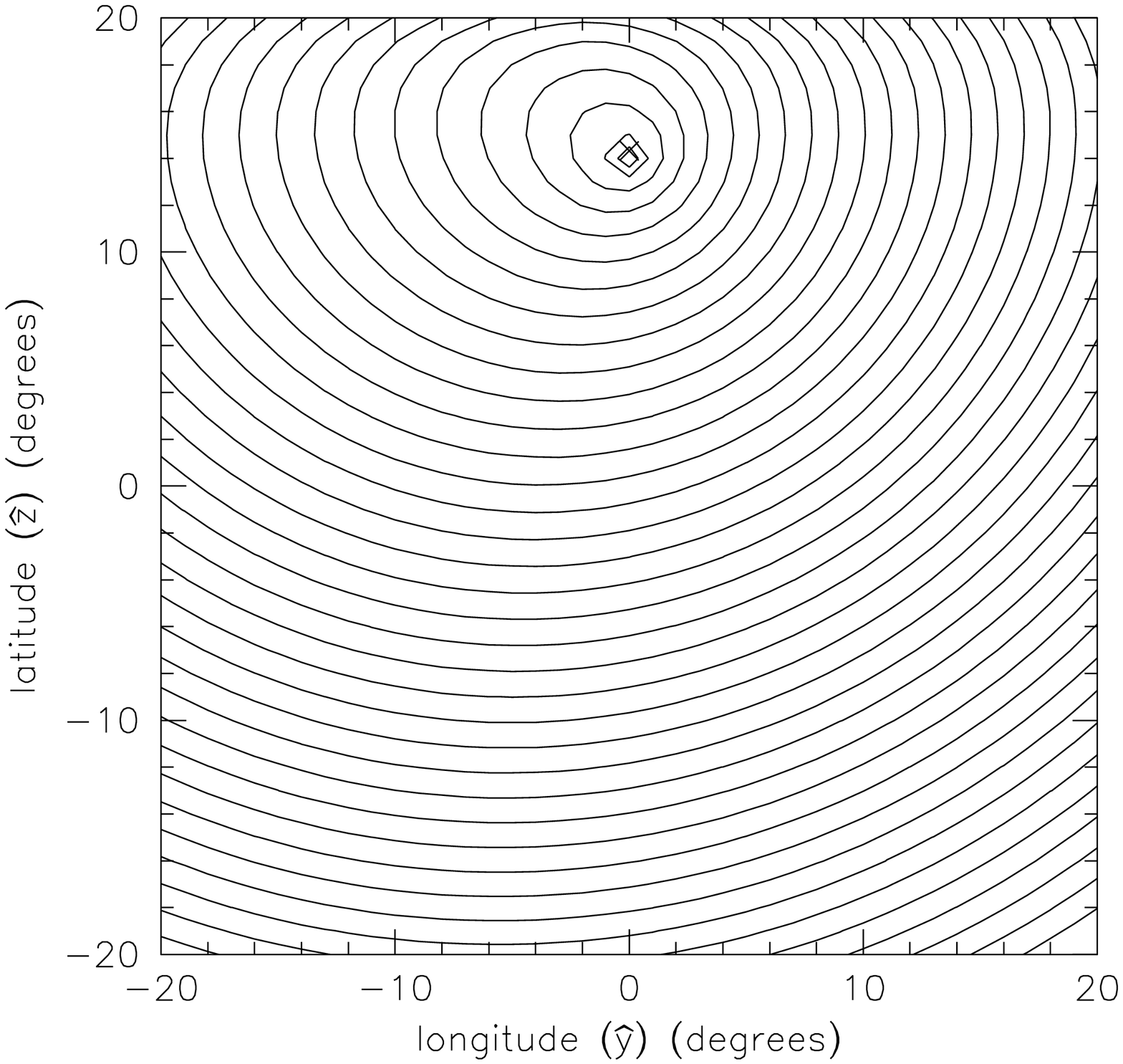,width=0.49\textwidth}
\epsfig{file=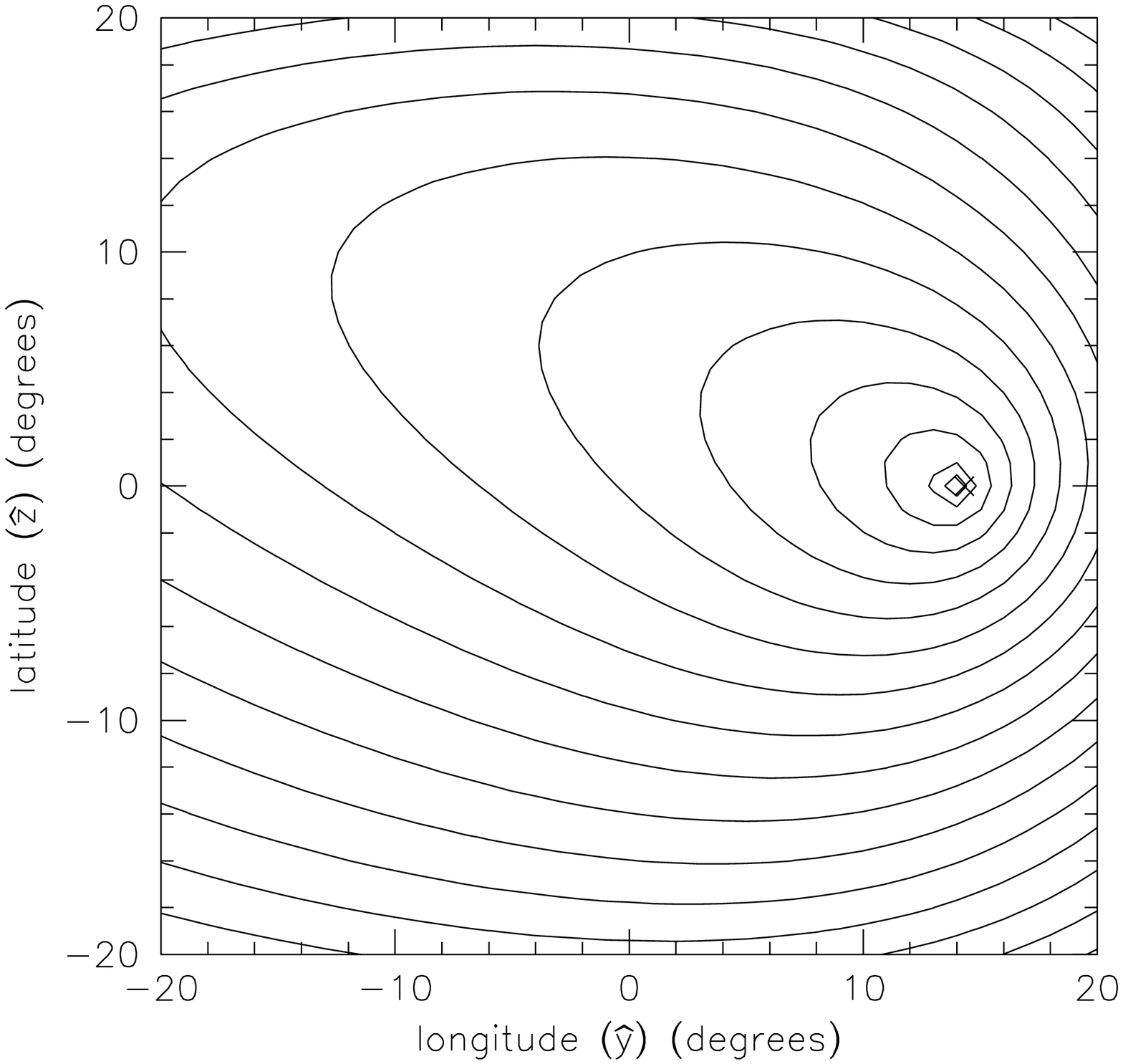,width=0.49\textwidth}
\caption{Total power in diffuse emission.  The normalization is arbitrary; this
figure is meant to illustrate the spatial extent of the signal.  The contours
are separated by 0.25 decade in power.  Four satellite geometries are
illustrated, each with a velocity of 300 km s$^{-1}$.  The annihilation
spectrum is that of the left panel of Fig.~\ref{fig:inputspec}.  In each case
the current angular position is indicated with a cross.  {\em Top left:}
$(x,y,z)=(8,0,1)$ kpc, $\hat{v}=(0,5/13,12/13)$. In this case the clump has
crossed more than half of the diffusion zone.  {\em Top right:}
$(x,y,z)=(12,0,4)$ kpc, $\hat{v}=(1,-1,1)/\sqrt{3}$. This illustrates the case
where the clump has passed completely through the diffusion zone, leaving
behind a wake that will dissipate shortly.  A hint of the edge of the diffusion
zone can be discerned at latitude $\approx 14^\circ$.  {\em Bottom left:}
$(x,y,z)=(8,0,2)$ kpc, $\hat{v}=(0,1,-1)/\sqrt{2}$.  In this case the clump has
recently entered the diffusion zone, thus the wake has not had much chance to
form, and the diffuse flux drops very rapidly. {\em Bottom right:}
$(x,y,z)=(8,2,0)$ kpc, $\hat{v}=(0,12,-5)/13$.  This final panel depicts a
clump moving nearly parallel to the galactic plane, thus the time spent inside
the diffusion zone is relatively long, and the tail becomes more extended.}
\label{fig:totalpower}
\end{figure}

The energy loss time $\tau$ represents all energy loss mechanisms: inverse
Compton scattering from both the CMB and starlight, and synchrotron emission.
The total power emitted in any one of these is inversely proportional to the
timescale for the individual process: $\tau_\star\approx 2\tau$, $\tau_{\rm
CMB}\approx 4\tau$, and $\tau_{\rm sync}\approx 4\tau$.  Integrating in energy,
\begin{equation}
P_X=\frac{1}{4\pi E_0\tau_X}\int dE\,E^2\frac{d\sigma}{dE},
\end{equation}
and we note this has units power per area per solid angle.  In
Fig.~\ref{fig:totalpower} we illustrate contours of the total power in diffuse
emission for several geometries of satellites.  Finally, we note that the
density of charged particles at the position of the Earth coming from the clump
is typically unobservably small.

\subsection{Diffuse emission: inverse Compton spectrum}

For electrons of Lorentz factor $\gamma=E/m_e$ scattering from a blackbody
spectrum of photons with $h\nu_T=kT$, the photon spectrum is given by the usual
inverse Compton formula \cite{longair}, integrated over the blackbody spectrum
\begin{equation}
\frac{1}{P_X}\frac{dP_X}{dx}=\frac{9}{128\zeta(3)}\,x^2\int_0^4
\frac{dy}{y}\frac{2\ln(y/4)+1+4/y-y/2}{\exp(x/y)-1},
\end{equation}
with $x=\nu/(\nu_T\gamma^2)$.  It is easy to show that this expression
integrates to unity.  This formula is inherently non-relativistic, implying
that $\gamma h\nu_T \ll m_e c^2$.  For the CMB, this is easily satisfied, as
$h\nu_T=0.2348$ meV for a 2.725~K blackbody spectrum.  For starlight, we assume
a blackbody spectrum with $T=3800$~K \cite{icstar}, thus $h\nu_T=0.33$ eV.  The
non-relativistic condition is satisfied for electron energies $E<$ TeV.  This
is true for most if not all of the annihilation products we will consider.  We
use $P_X=E^2/(E_0\tau_X)$ to find the spectrum from a single particle (in power
per frequency),
\begin{equation}
\frac{dP_X}{d\nu}=\frac{9}{128\zeta(3)}\,\frac{m_e^2}{E_0\tau_X\nu_T}\,
x^2\int_0^4\frac{dy}{y}\frac{2\ln(y/4)+1+4/y-y/2}{\exp(x/y)-1}.
\end{equation}
The spectrum from the full column of particles is then simply (in power per
area per frequency)
\begin{equation}
\frac{dP_X}{d\nu}=\frac{m_e^2}{4\pi E_0\tau_X\nu_T}\int dE\,
\frac{d\sigma}{dE}\,\left[\frac{1}{P_X}\frac{dP_X}{dx}
\left(x=\frac{\nu m_e^2}{\nu_T E^2}\right)\right].
\label{eq:fullspectrum}
\end{equation}

Detecting a diffuse signal in gamma rays will necessarily be hampered by at
least the extragalactic background.  The EGRET satellite measured this to be
\cite{egretbg}
\begin{equation}
E\,\frac{d\Phi}{dE}=(3.30\pm0.15)\times 10^{-6}
\left(\frac{E}{0.451\;\rm GeV}\right)^{-1.10\pm0.03} \rm
cm^{-2}\; s^{-1}\; sr^{-1}.
\end{equation}

\subsection{Diffuse emission: synchrotron spectrum}

\begin{figure}
\epsfig{file=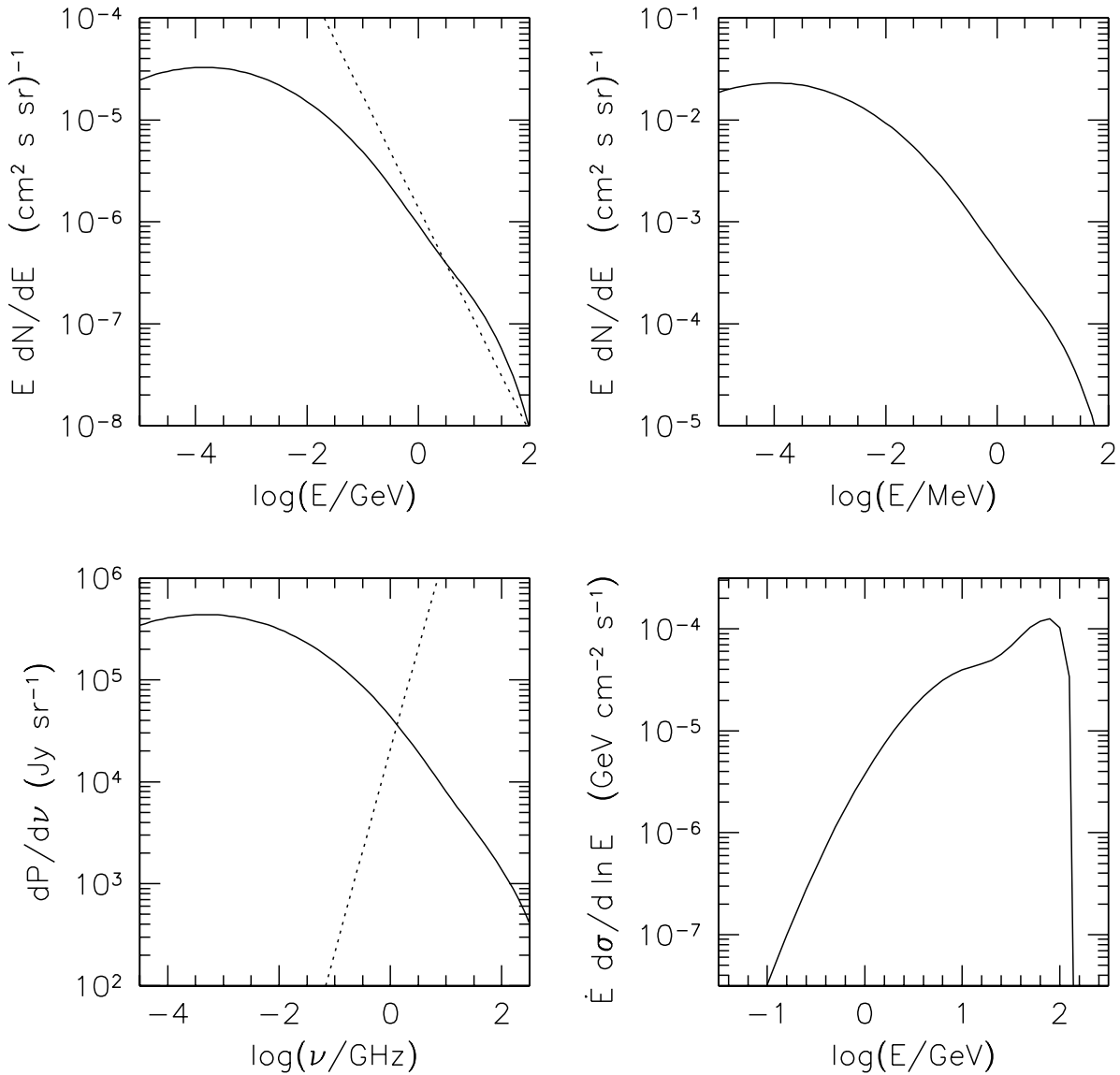,width=0.95\textwidth}
\caption{Spectra of diffuse emission in inverse Compton and synchrotron
radiation.  The clump is located at $(x,y,z)=(8,0,1)$ kpc, with a velocity of
300 km s$^{-1}$ in the $\hat{v}=(0,5/13,12/13)$ direction, as in the top left
panel of Fig.~\ref{fig:totalpower}.  The annihilation spectrum is that of the
left panel of Fig.~\ref{fig:inputspec}.  The line of sight illustrated trails
the current location of the clump by 5 degrees in the $-\hat{v}$ direction.
The total annihilation rate is 10$^{39}$ s$^{-1}$.  {\em Top left:} Spectrum of
inverse Compton emission from interactions with starlight, assuming an energy
density of 0.6 eV cm$^{-3}$ with a 3800K blackbody spectrum.  The dotted curve
is the diffuse gamma ray background.  {\em Top right:} Inverse Compton on the
CMB.  {\em Bottom left:} Synchrotron emission assuming a 3 $\mu$G magnetic
field.  The dotted curve is the CMB.  {\em Bottom right:} Column depth of
particles as a function of energy, weighted by the energy loss rate.  It is
clear that most of the diffuse power is radiated by relatively high energy
particles.}
\label{fig:diffspec}
\end{figure}

The total power in synchrotron radiation is similar to that in inverse Compton
emission in that both are proportional to $E^2$.  The spectra are different
however.
\begin{equation}
\frac{1}{P_{\rm sync}}\frac{dP_{\rm sync}}{dx}=
\frac{27\sqrt{3}}{32\pi}\,x\int_0^\pi d\alpha\,\sin\alpha
\int_{x/\sin\alpha}^\infty dy\,K_{5/3}(y),
\end{equation}
and now $x=\nu/(\nu_B\gamma^2)$, with $\nu_B=3eB/(4\pi m_ec)=12.6\,(B/3\mu\rm
G)$ Hz.  This is just the usual synchrotron formula, isotropically integrated
over the angle the particle trajectory makes with the magnetic field.  It is
again easy to show that this expression integrates to unity.  The spectrum from
the full column of particles is then easily recovered from
Eq.~\ref{eq:fullspectrum}, replacing the $(dP/dx)/P$ with the synchrotron
formula, and replacing $\nu_T$ with $\nu_B$ (and we take $B=3\mu$G).  The chief
diffuse background for this synchrotron signal is the CMB, which peaks at 160
GHz.  At 1 GHz, the power is $2.09\times 10^4$ Jy sr$^{-1}$.  In
Fig.~\ref{fig:diffspec} we plot the spectra of inverse Compton and synchrotron
emission from a line of sight that is 5 degrees behind the current position of
a clump.  We will neglect complications such as synchrotron self-absorption and
synchrotron self-Compton processes as we calculate the photon density to be
much smaller than in previous treatments \cite{blasi} due to the diffusion of
the source electrons.

\section{Galactic halo substructure}

Many authors have discussed the distribution of substructure in galactic halos.
There are two important issues relating to the detectability of annihilations
in these structures, both their mass distribution and their density profile.
Again we stress that we are interested only in those satellites within the
diffusion zone.  As the diffusion zone is fairly extended, there should be a
significant number of such satellites: even out to a radius of 30 kpc, the
diffusion zone is roughly 15\% of the total volume of the galactic halo.

\subsection{Density profiles}
We first discuss the density profiles of the satellite halos that we hope to
detect.  Numerical N-body simulations find that structures at all scales
diverge as a power law at small radii.  In particular, the popular NFW profile
\cite{nfw} is given by
\begin{equation}
\rho(r)=\rho_0\left(\frac{a}{r}\right)\frac{a^2}{(a+r)^2}.
\end{equation}
These profiles can be normalized according to simulations, based on the virial
mass and radius, where the density is roughly 340 times the cosmological
density (for the concordance model) $\rho_c=3\Omega H_0^2/8\pi
G\approx1.4\times10^{-6}\,\rm GeV\;cm^{-3}$.  The relevant parameter is the
concentration $C=r_{\rm vir}/a$, with $r_{\rm vir}=260(M_{\rm
vir}/10^{12}\,M_\odot)^{1/3}\,{\rm kpc}$ \cite{concentration}, finding
\begin{eqnarray}
C&\approx&107\left(\frac{M_{\rm vir}}{M_\odot}\right)^{-0.084},\\
\rho_0&=&340\, C(1+C)^2\rho_c,\\
M_{\rm vir}&=&4\pi\rho_0\left(\frac{r_{\rm vir}}{C}\right)^3
\left[\ln\left(1+c\right)-\frac{C}{1+C}\right].
\end{eqnarray}
The specific annihilation rate is then given by
\begin{equation}
\frac{\Gamma}{M_{\rm vir}}=\frac{\sigv\rho_0}{3m^2}
\frac{1-(1+C)^{-3}}{\ln(1+C)-C/(1+C)}\rightarrow
\frac{\sigv\rho_0}{3m^2}
\frac{1}{\ln C -1}\;\;(C\gg1).
\end{equation}
The Moore profile has a steeper central cusp with a divergent total flux
\cite{moore}.  Thus a minimum radius $r_{\rm min}$ needs to be defined to
regularize.  The profile is given by
\begin{eqnarray}
\rho(r)&=&\rho_0\left(\frac{a}{r}\right)^{1.5}\frac{1}{[1+(r/a)^{1.5}]},\\
\frac{\Gamma}{M_{\rm vir}}&\rightarrow&\frac{\sigv\rho_0}{m^2}
\frac{\ln(a/r_{\rm min})}{\ln C}.
\end{eqnarray}
For cusps $\rho\propto r^{-\alpha}$ steeper than $\alpha=1.5$, we find the
behavior (with the proper limit as $\alpha\rightarrow 1.5$)
\begin{equation}
\frac{\Gamma}{M_{\rm vir}}\rightarrow\frac{\sigv\rho_0}{m^2}
\frac{(a/r_{\rm min})^{2\alpha-3}-1}{(2\alpha-3)\ln C}.
\end{equation}

\subsection{Mass distribution of subhalos}

Simulations indicate that the mass distribution of subhalos is a power law,
given by
\begin{equation}
N\propto\left(\frac{M}{M_{\rm vir}}\right)^{-\alpha},
\end{equation}
with a normalization that about 500 clumps with masses larger than
$10^8\,M_\odot$ are found in a halo like that of the Milky Way
\cite{clumpspectrum}.  This means that there will be only a handful of clumps
more massive than $10^{10}\,M_\odot$.  At $M_{\rm vir}=10^9\,M_\odot$, the
concentration is $C\approx 19$.  The specific annihilation rate for a clump is
easily determined, using $\sigv\rho_c/m^2=5\times
10^{24}\,(\sigv/10^{-27}\,{\rm cm^3\; s^{-1}})\,(m/{\rm
GeV})^{-2}\,M_\odot^{-1}\,\rm s^{-1}$.  Using $\rho_0\approx 340\rho_c C^3$ for
any of these halo models, we can find a simple expression for the total
annihilation rate in a ``large'' clump, defined as $M_{\rm vir}=10^9\,M_\odot$,
$r_{\rm vir}=26$ kpc (and $a=1.4$ kpc):
\begin{eqnarray}
\Gamma_{\rm NFW}&=&2\times 10^{39}\,\frac{\sigv_{27}}{m^2_{\rm GeV}}\;\rm
s^{-1},\\
\Gamma_{\alpha}&=&4\times 10^{39}\,\frac{(a/r_{\rm min})^{2\alpha-3}-1}
{2\alpha-3}\,\frac{\sigv_{27}}{m^2_{\rm GeV}}\;\rm s^{-1}.
\end{eqnarray}
In supersymmetric models, $\sigv_{27}/m^2_{\rm GeV}<10^{-2}$.  The Moore
profile has a logarithmically diverging flux, but the enhancement factor is
less than 40 if annihilations empty the central cusp, and much less than that
for other mechanisms.  For example, if interactions with a central black hole
sweep out the central region (of order 1 pc), the enhancement is less than 10.
This may be the case for the full galactic halo, but for dark clumps without
stars, there may be no central black hole.  Steeper profiles can in principle
have much larger enhancement factors, though with a cutoff of 1 pc, the
enhancement is less than of order 1000 even for $\alpha=2$.  In the next
section, we will see that annihilation rates of order $10^{38}\, \rm s^{-1}$
are needed for point sources to be confidently seen in diffuse emission, thus
$\alpha\ge1.5$ is probably required.  Coincidentally, steeper profiles look
more like point sources, so clumps with high enough annihilation rates will be
easy to model from this perspective.  However, the NFW profile is not too far
from detectability for a 10$^9 M_\odot$ clump, if the annihilation cross
section is the maximum allowed.  Some clumping of the halo might tip the
balance and make even this profile a candidate for detection in inverse Compton
emission.  Here we note that the effect that clumpiness in galactic halos has
on dark matter detection has been discussed at length,
e.g.~Ref.\cite{clumping}.

\section{Unidentified EGRET sources}

The EGRET gamma ray survey discovered a large number of point sources, a large
fraction of which remain unidentified \cite{egretpoint}.  These sources
typically have fluxes of order
\begin{equation}
E\,\frac{d\Phi}{dE}\sim 10^{-8}\left(\frac{E}{\rm
GeV}\right)^{-1}\, {\rm cm^{-2}\;s^{-1}}.
\end{equation}
We investigate the possibility that some of these might be Milky Way satellites
undergoing annihilations at their centers.  The annihilation scenario requires
that there is no other concentrated emission than the gamma rays.  Other
astrophysical sources of just about any type are expected to have emission at
other wavelengths as well.  These sources (identified or not) are not measured
below about 100 MeV, thus the turnover expected for $\pi^0$ photons at
$m_{\pi^0}/2=67.5$ MeV could not have been observed.  However, the spectrum of
an annihilation source should be closer to flat around 100 MeV, as seen from
Fig.~\ref{fig:inputspec}, thus the EGRET point sources are typically not very
good candidates for annihilation radiation.  However, they do set the scale of
possible annihilation sources quite well.

The yield of gamma rays predominantly from $\pi^0$ decays in the hadronization
of annihilation products is typically $dN/d\,\ln\,E\sim2$ annihilation$^{-1}$
at energies of $0.1$ GeV.  This means the inferred rate of annihilations for
these sources is $5\times10^{-8}$ cm$^{-2}$ s$^{-1}$.  For a clump at a
distance of 8 kpc, the inferred annihilation rate is $4\times 10^{38}$
s$^{-1}$.  As we have seen in the previous section, this rate likely requires a
central density cusp at least as steep as $r^{-1.5}$.

\begin{figure}
\epsfig{file=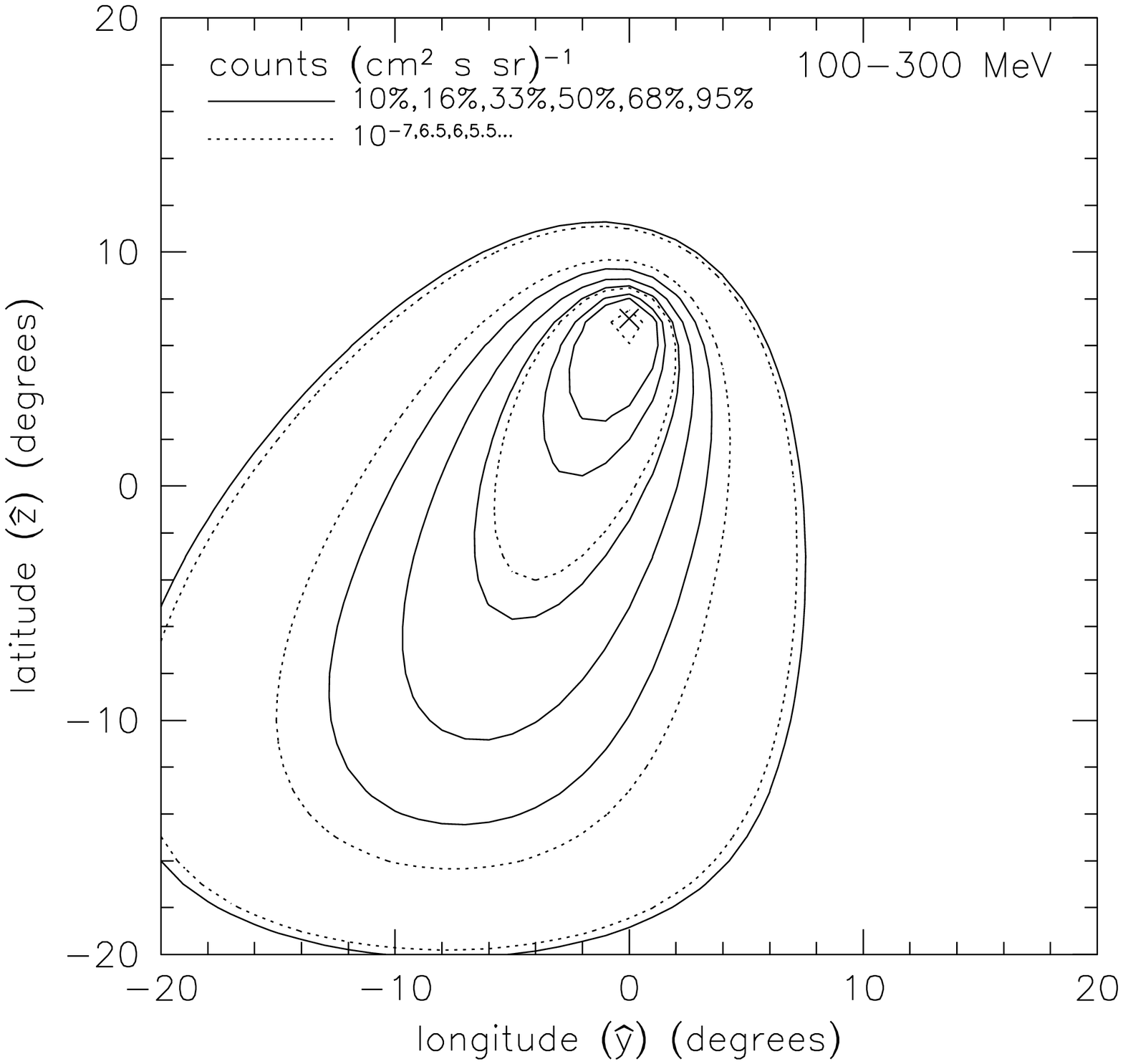,width=0.49\textwidth}
\epsfig{file=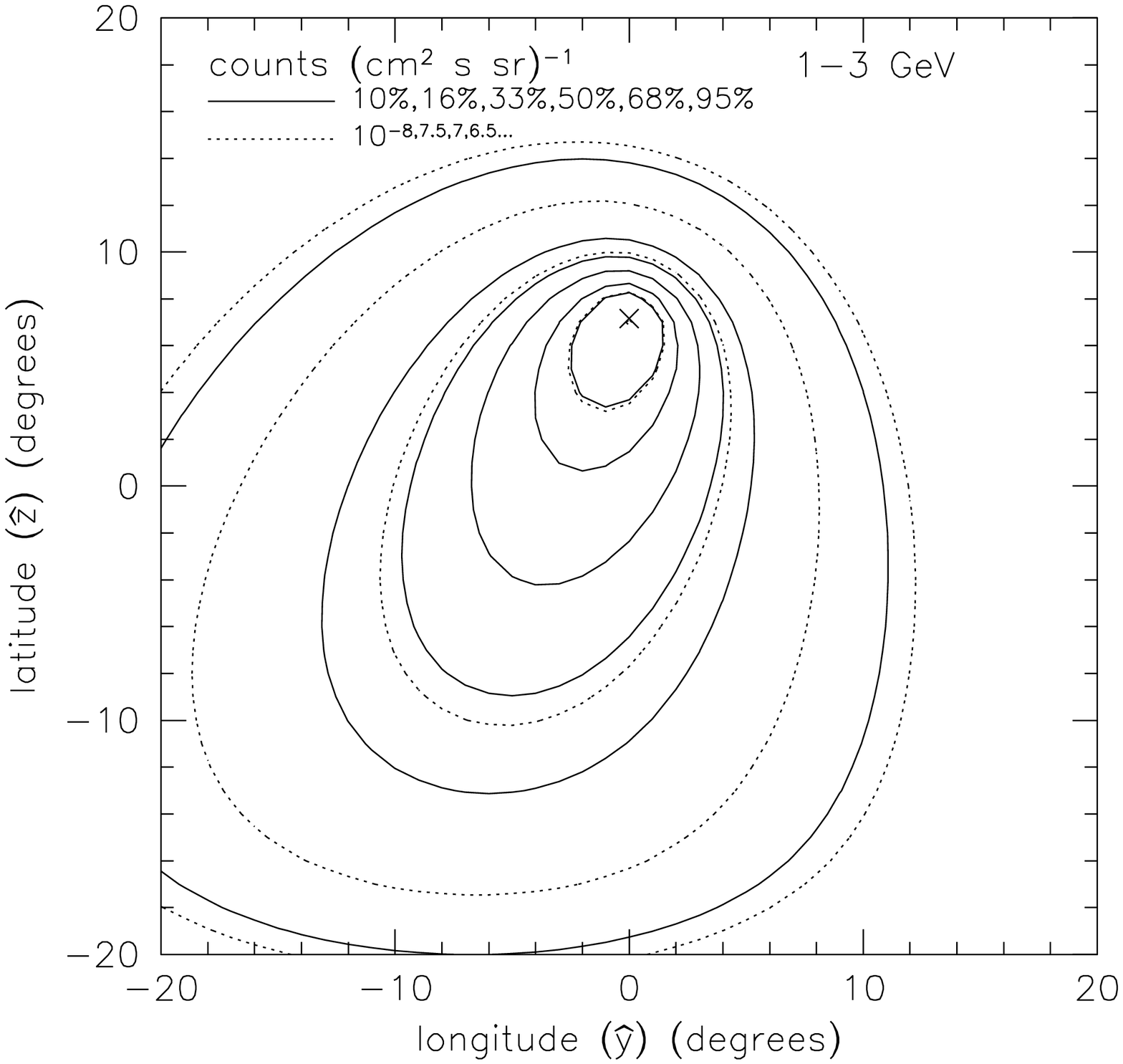,width=0.49\textwidth}
\epsfig{file=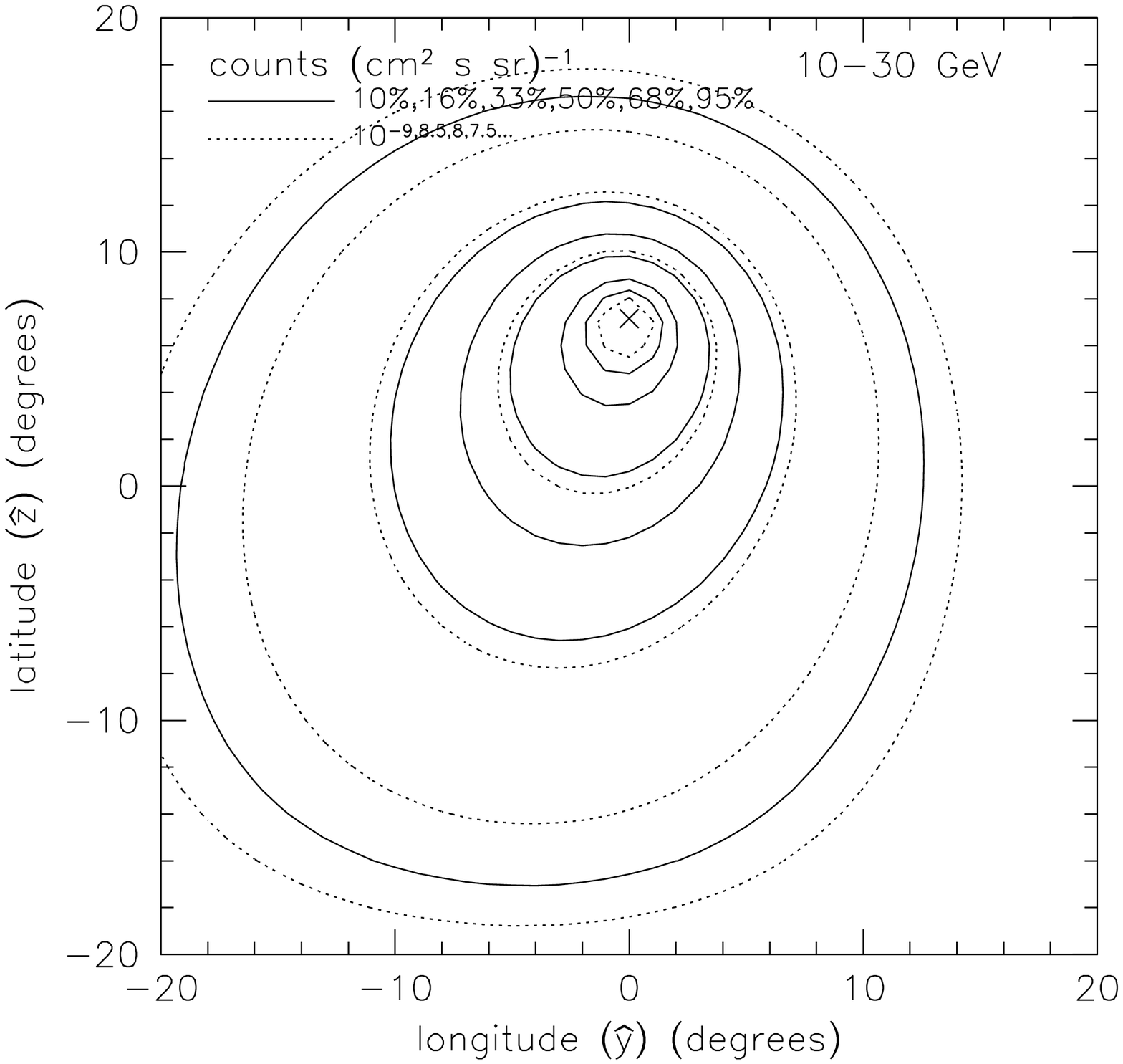,width=0.49\textwidth}
\epsfig{file=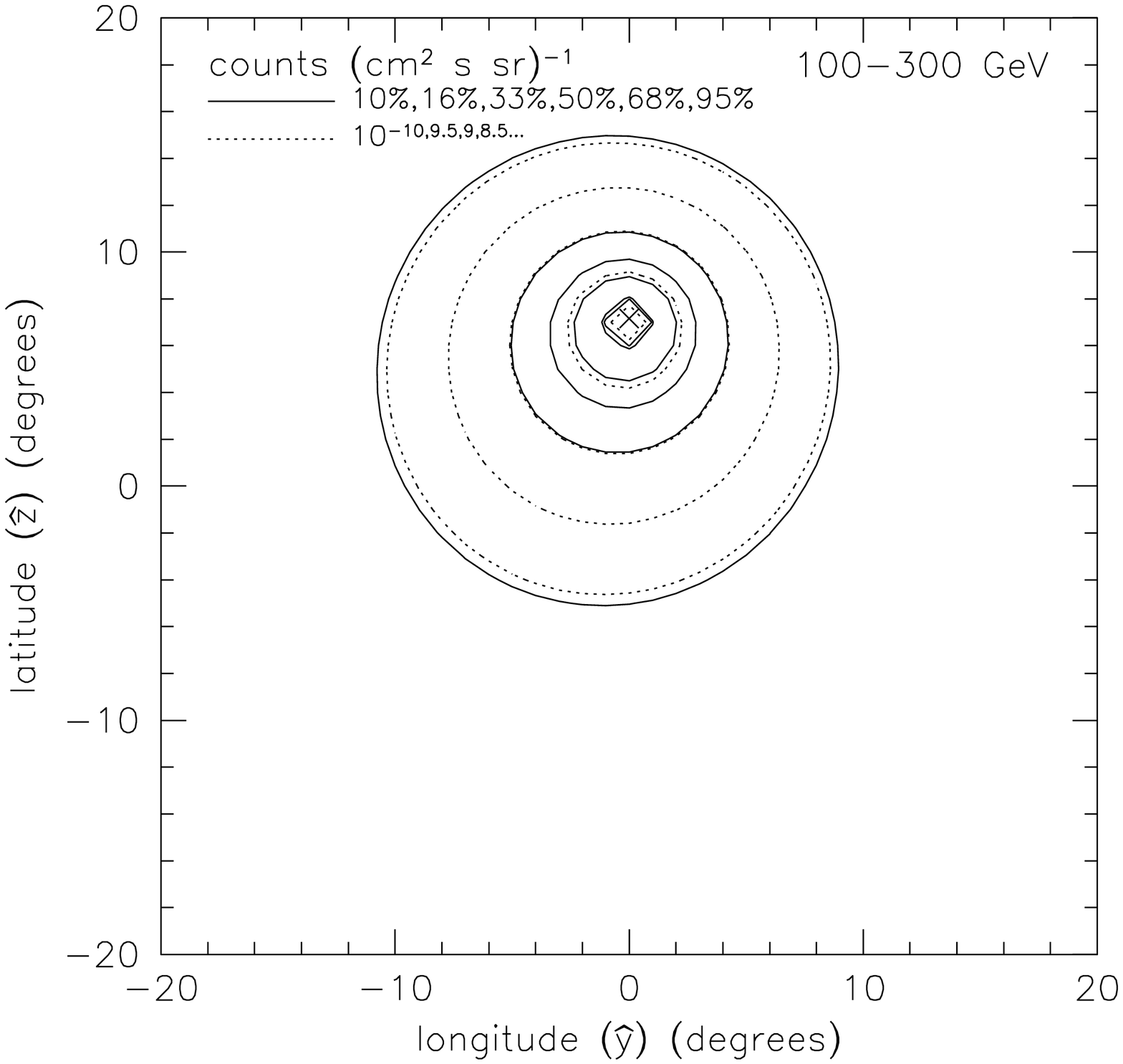,width=0.49\textwidth}
\caption{Contours of integrated diffuse flux in four energy bands.  The clump
is located at $(x,y,z)=(8,0,1)$ kpc, with a velocity of 300 km s$^{-1}$ in the
$\hat{v}=(0,5/13,12/13)$ direction, as in the top left panel of
Fig.~\ref{fig:totalpower}.  An annihilation source of $4\times 10^{38}$
s$^{-1}$ is assumed, with the spectrum of the left panel of
Fig.~\ref{fig:inputspec}.  Two sets of contours are plotted: the solid ones
indicate the fraction of diffuse flux contained for 10\%, 16\%, 33\%, 50\%,
68\%, and 95\%, while the dotted ones are in units of photons cm$^{-2}$
s$^{-1}$ sr$^{-1}$, separated by 0.5 decade in flux.  The flux is very
asymmetric at low energy, becoming nearly round at the highest energies.}
\label{fig:mapspec}
\end{figure}

We ask the following question: can GLAST confirm or rule out the hypothesis
that some of the unidentified EGRET point sources are the annihilating cores of
very steep profile dark matter subhalos?  We argue that the answer is yes, for
two reasons.  First, GLAST will have improved performance an energies below 100
MeV, down to a threshold of 20 MeV.  Certainly, the turnover in spectrum at
about 70 MeV that should be present in an annihilation source should be
visible.  Second, the diffuse inverse Compton emission associated with a few
times 10$^{38}$ annihilations per second, covering hundreds of square degrees,
should yield both a significant fraction of the background and a statistically
significant detection.  GLAST will have an enormously improved exposure over
EGRET: at 1 GeV and above, the expectation is $4.5\times 10^{10}$ cm$^2$ s to
any point on the sky, per year \cite{glastexposure}.  In
Fig.~\ref{fig:mapspec}, we illustrate the integrated flux from inverse Compton
on starlight in four energy intervals due to an annihilation source of $4\times
10^{38}$ s$^{-1}$ with the spectrum of the left panel of
Fig.~\ref{fig:inputspec}.  We find that such a source should be easy for GLAST
to detect, in fact a source ten times less active, with $4\times10^{37}$
annihilations per second should be detectable.  These results are summarized in
Table.~\ref{tab:egretglast}.

\begin{table}
\begin{ruledtabular}
\begin{tabular}{ccccccc}
Energy & Size (68\%) & S / BG & BG (EGRET) & BG (GLAST) &
S/N (EGRET) & S/N (GLAST) \\
(GeV) & (sr) & & (counts) & (counts)\\
\hline 
0.03 -- 0.1 & 0.078 & 0.551 & 250. & 5,630 & 8.72 & 41.4\\
0.1 -- 0.3 & 0.084 & 0.961 & 75.2 & 4,230 & 8.33 & 62.5\\
0.3 -- 1 & 0.093 & 1.447 & 23.6 & 2,650 & 7.03 & 74.6\\
1 -- 3 & 0.102 & 2.098 & 7.25 & 1,630 & 5.65 & 84.8\\
3 -- 10 & 0.102 & 3.209 & 2.06 & 463. & 4.60 & 69.0\\
10 -- 30 & 0.076 & 5.322 & 0.43 & 97.1 & 3.49 & 52.4\\
30 -- 100 & 0.042 & 8.019 & 0.07 & 15.1 & 2.08 & 31.2\\
100 -- 300 & 0.021 & 6.562 & 0.01 & 2.10 & 0.63 & 9.50
\end{tabular}
\end{ruledtabular}
\caption{Detectability of diffuse inverse Compton emission in EGRET and GLAST.
An annihilation source of $4\times 10^{38}$ s$^{-1}$ is assumed, with the
spectrum of the left panel of Fig.~\ref{fig:inputspec}.  The EGRET exposure is
assumed to be 10$^9$ cm$^2$ s.  For GLAST, we assume a 5 year mission, with a
total exposure of $2.25\times 10^{11}$ cm$^2$ s for energies above 1 GeV.  In
the energy bins below 1 GeV, we assume exposures of 50\% (0.3--1 GeV), 25\%
(0.1--0.3 GeV), and 10\% (0.03--0.1 GeV) of this value \cite{glastexposure}.
The size column indicates the 68\% containment region.  The signal / background
column represents the total number of photons, inside the 68\% containment
region, compared with the extragalactic background.  For a convincing
detection, we require that the signal be an appreciable fraction of the
background, as the anisotropies in the background are in principle
considerable.  We of course also require a significant signal to noise ratio.
Lastly, we require a significant total number of photons.  Based on these rough
criteria, the detectability of the diffuse emission by EGRET seems marginal,
though GLAST should be able to detect it easily.  If the annihilation rate were
ten times smaller, GLAST should be able to find the signal as well.  Note that
``3'' in the energy bands means $\sqrt{10}$, i.e.\ the energy bands cover 0.5
decade each.}
\label{tab:egretglast}
\end{table}

\section{Diffuse synchrotron emission}

An annihilation rate of $10^{39}$ s$^{-1}$ produces a significant amount of
diffuse synchrotron emission.  Previous authors have discussed this in the
context of clumps \cite{blasi}, neglecting the effects of diffusion.  Their
calculation essentially assumed a point source of synchrotron radiation
(actually $10'\times10'\approx 10^{-5}$ sr), with the same total power as we
find.  We calculate that the diffusion of the charged particles spreads out the
synchrotron signal over roughly $0.1$ sr, a huge area compared with the
previous assumption.  For this clump, the synchrotron emission exceeds the CMB
only below 1 GHz.  In principle, with spectral information the synchrotron and
CMB can be untangled, but as the signal lies near the galactic plane, this may
not be feasible.

\section{The galactic center source}

\begin{figure}
\epsfig{file=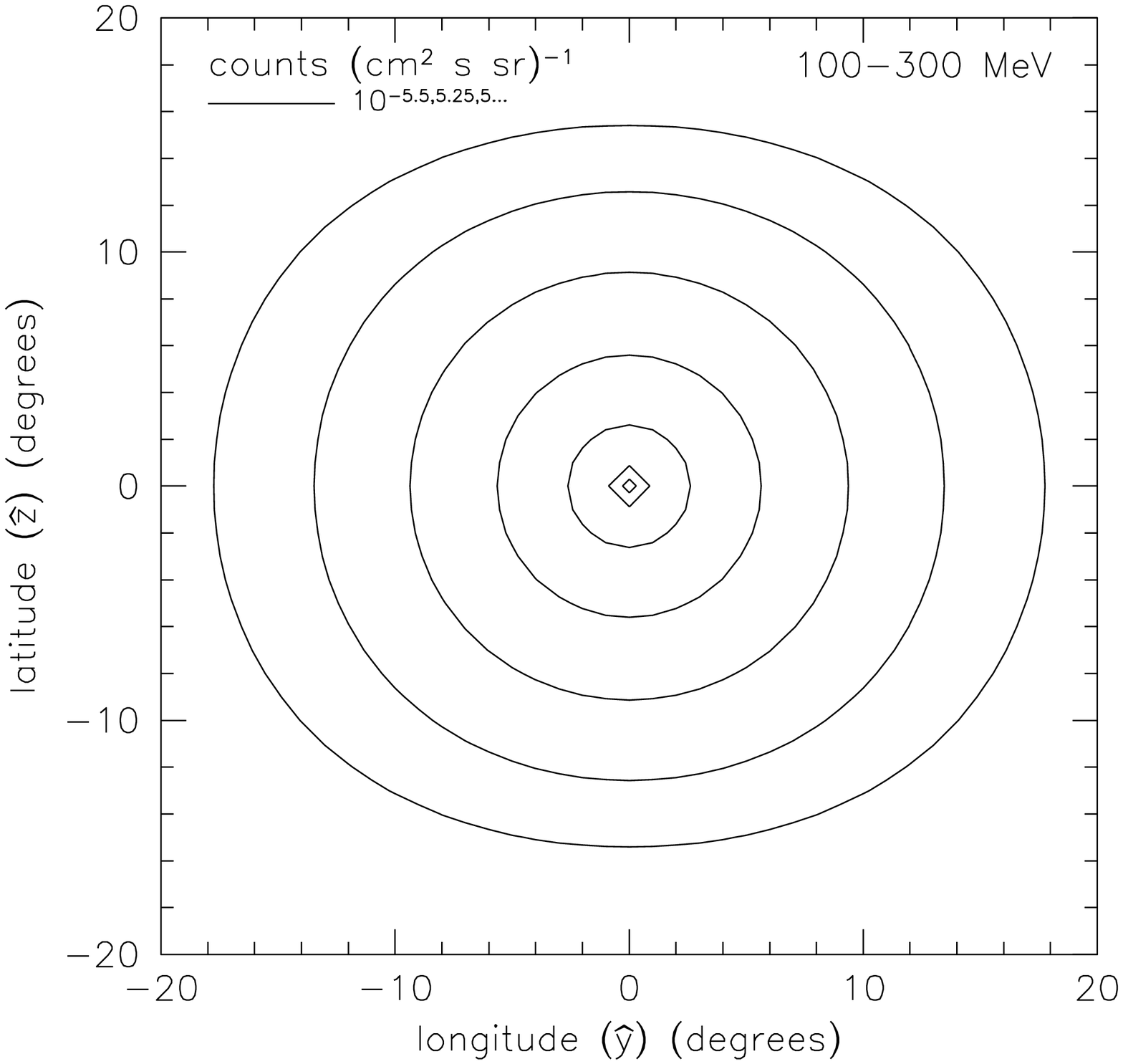,width=0.49\textwidth}
\epsfig{file=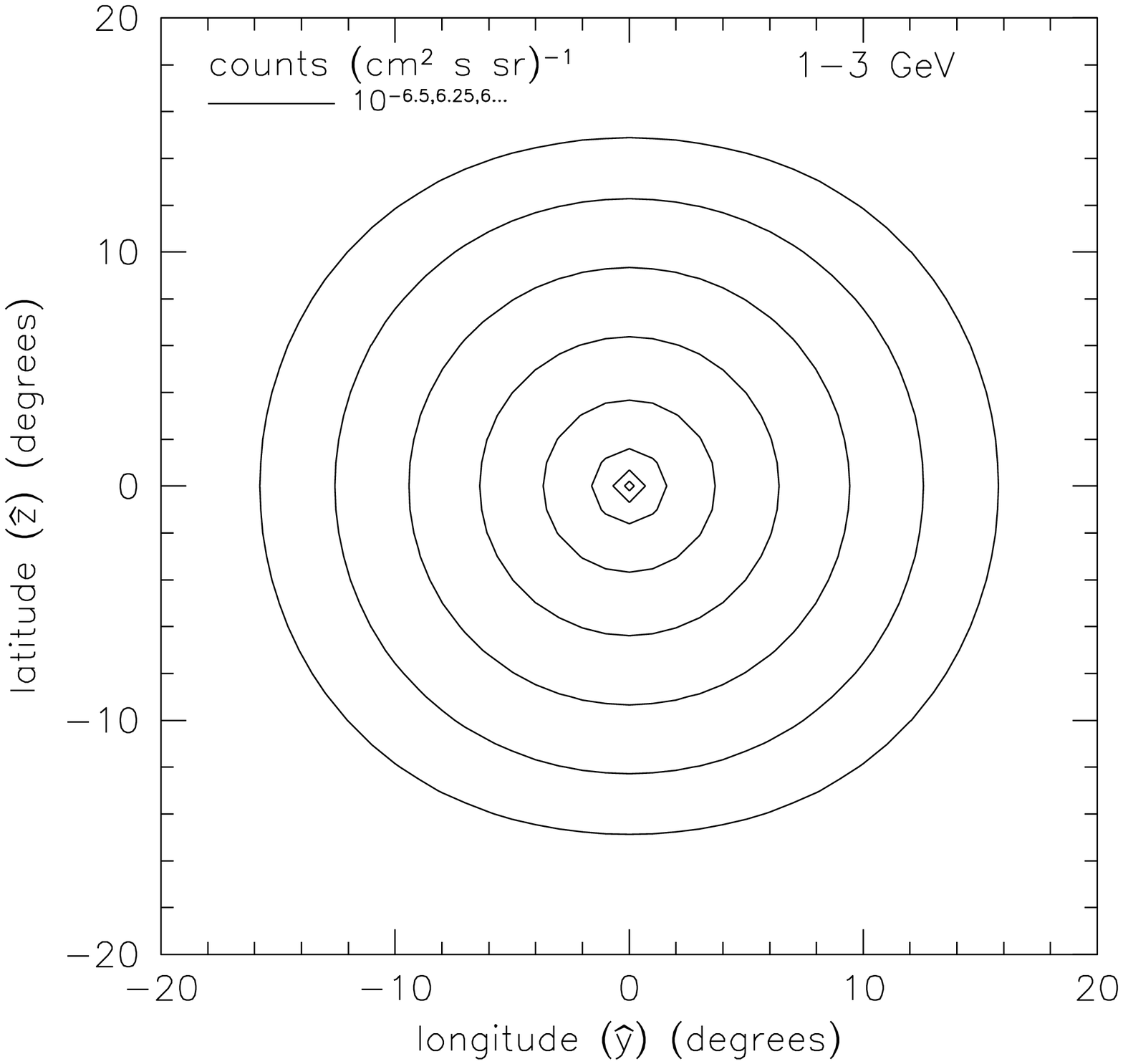,width=0.49\textwidth}
\caption{Contours of integrated diffuse flux in two energy bands associated
with the galactic center source.  The source is assumed to be fixed at
$(x,y,z)=(8.5,0,0)$ kpc.  An annihilation source of $4\times 10^{39}$ s$^{-1}$
is assumed, with the spectrum of the left panel of Fig.~\ref{fig:inputspec}.
The contours are in units of photons cm$^{-2}$ s$^{-1}$ sr$^{-1}$, separated by
0.25 decade in flux.}
\label{fig:mapspecgc}
\end{figure}

The galactic center is known to be a bright source of gamma rays, with a broken
power law spectrum measured by EGRET to be \cite{egretgc}
\begin{eqnarray}
E\,\frac{d\Phi}{dE}&=&(4.2\pm0.02)\times10^{-7}\left(\frac{E}{1.9\;\rm
GeV}\right)^{-\alpha}\rm cm^{-2}\;s^{-1},\\
\alpha &=&0.3\pm0.03\;\;(E<1.9\;\rm GeV),\\
\alpha &=&2.1\pm0.20\;\;(E>1.9\;\rm GeV).
\end{eqnarray}
We briefly discuss the possibility that this is an annihilation source.  At 100
MeV, this spectrum corresponds to $5\times10^{-7}$ annihilations cm$^{-2}$
s$^{-1}$.  Assuming the galactic center source is at a distance of 8.5 kpc, the
inferred annihilation rate is $4\times10^{39}$ s$^{-1}$.  This is a factor of
ten brighter than the bright clumps discussed previously.  In
Fig.~\ref{fig:mapspecgc} we illustrate the inverse Compton emission that would
be associated with the galactic center source.  We find that the emission
corresponds to a small fraction of the diffuse emission near the galactic
center \cite{egretgcdiffuse}, and furthermore, it falls with distance from the
center much more quickly than the observed halo (for an analysis of the
gamma-ray halo see Ref.~\cite{SMRgc}).  If the galactic center region were
better understood, it might be possible to uncover such a signature.

\section{Discussion and conclusions}

We have calculated the spectrum and spatial extent of diffuse emission from the
charged particle products of dark matter annihilations.  In addition to the
synchrotron emission discussed previously, we have studied the inverse Compton
radiation, primarily on starlight photons.  We have focused on galactic
satellites that are currently within the diffusion zone, namely within a few
kpc of the stellar disk.  For satellites moving with typical galactic halo
velocities of 300 km s$^{-1}$, the crossing time of the diffusion zone is of
the same order as the diffusion time, thus an inherently time-dependent
treatment is required.

For annihilation sources, e.g.\ galactic satellites at typical distances of 10
kpc, the diffuse emission in both inverse Compton and synchrotron extends over
roughly 300 square degrees.  We have shown that at least in terms of the number
of photons, the diffuse inverse Compton emission might be detectable by GLAST,
assuming bright enough annihilation sources.  The spatial extent of the
emission makes its detection problematic of course.  GLAST will certainly
detect a significant number of point sources in a region of this size.  In a
future work we will study in detail the feasibility of separating these
signals.

As mentioned previously, these results are fairly generic, and do not depend
strongly on the particle physics model.  As we are concerned with the electrons
and positrons, we do not even require that the dark matter have hadronic
interactions.  Leptonically interacting dark matter
\cite{krauss,baltzbergstrom} would still provide photons and electrons, albeit
by different processes and with different spectra.  Such photon sources would
be even harder to reconcile with the EGRET point sources, but annihilation
sources below the EGRET detection limit may be detectable by GLAST in any case.

\begin{acknowledgments}
We thank H.~Tajima and P.~Gondolo for interesting conversations.  This work was
supported in part by the U.S. Department of Energy under contract number
DE-AC03-76SF00515.
\end{acknowledgments}

%%%%%%%%%%%%%%%%%%%%%%%%%%%%%%%%%%%%%%%%%%%%%%%%%%%%%%%%%%%%%%%%%%%%%
%       REFERENCES
%%%%%%%%%%%%%%%%%%%%%%%%%%%%%%%%%%%%%%%%%%%%%%%%%%%%%%%%%%%%%%%%%%%%%

\end{document}